\documentclass[aps,twocolumn,amsmath,amssymb,floatfix,showpacs,
groupedaddress]{revtex4}

\usepackage{bm,amsmath,amssymb,amsfonts}

\usepackage[dvips]{graphicx}
\usepackage{array}

\usepackage{color}
\definecolor{mycolor}{rgb}{0.30,0.50,0.20}
\begin{document}

\title{\textcolor{mycolor}{Engineering slow light and mode crossover in a fractal-kagom\'{e}
waveguide network}} 

\author{Atanu Nandy}
\email{atanunandy1989@gmail.com}

\author{Arunava Chakrabarti}
\email{arunava_chakrabarti@yahoo.co.in}

\affiliation{Department of Physics, University of Kalyani, Kalyani,
West Bengal-741235, India}

\begin{abstract}
We present an analytically exact scheme of unraveling 
a multitude of flat, dispersionless photonic bands 
in a kagom\'{e} waveguide strip where each elementary plaquette 
hosts a deterministic fractal geometry of arbitrary size. 
 The number of 
non-dispersive eigenmodes grows as higher and higher order fractal 
geometry is 
embedded in the kagom\'{e} motif. Such eigenmodes 
are found to be localized with finite support in the 
kagom\'{e} strip and exhibit a hierarchy of localization
areas. The onset of localization can, in principle, be delayed 
{\it in space} by an appropriate
choice of frequency of the incident wave. 
The length scale at which the onset of localization for each
mode occurs, can be tuned at will as prescribed here using a 
real space renormalization method. 
Conventional methods of extracting the non-dispersive modes in such 
geometrically frustrated lattices fail as a non-translationally invariant 
fractal decorates the {\it unit cells} in the transverse direction. The 
scheme presented here  
circumvents this difficulty, and thus may inspire the 
experimentalists to design similar fractal incorporated kagom\'{e} 
or Lieb class of lattices to observe a multifractal distribution of 
flat photonic bands.
\end{abstract}

\pacs{63.20.Pw, 42.25.Bs, 42.25.Kb, 42.82.Et}

\maketitle
\section{Introduction}
\label{intro}
The geometrically frustrated lattices (GFL) have shown growing interest over the 
past few years
not only for the emergence of exotic 
spin liquid, spin ice and charge and magnetic disordered phases 
in strongly correlated
systems ~\cite{mezza}-\cite{wu}, but also very recently for discerning 
the existence of flat non-dispersive
electronic states leading to remarkable band structures in a wide variety of physical systems 
~\cite{zheng}-\cite{derzhko2}, as well as in several artificial model lattices ~\cite{atanu2,atanu3}.
The topology of the underlying lattice and destructive quantum interference play a pivotal role 
in localizing single particle eigenstates even when there is no apparent disorder in the 
system. The eigenstates have  non-zero amplitudes either on specific atomic sites, 
or over clusters of sites which are {\it effectively separated} from their neighboring 
clusters by sites on which the amplitude of the wave function vanishes. 
This results in non-dispersive energy-wave vector relationship which makes the 
effective mass of an electron diverging, leading to an complete or 
extremely low mobilty of the particle
~\cite{atanu2,atanu3}. 

As is well known now, localization is common to classical waves 
as well, including light. The concept of {\it photonic band gap} (PBG) was initiated by 
Yablonovitch~\cite{yablo1,yablo2} along with the proposals from John~\cite{john} 
and Pendry and MacKinon~\cite{pendry} regarding the possibility of observing 
the Anderson localization of light.   
Recent direct experimental observation of localized photonic flat band states
~\cite{sebabrata,rodrigo} in 
Lieb photonic lattices and tight binding photonic bands in metallophotonic 
waveguide networks~\cite{endo} have enriched and stimulated research in photonics. 
Such `flat' photonic bands are intimately connected 
to the pathbreaking idea of engineering {\it slow light} with low group velocity 
which opens up the possibility of a `spatial compression of light energy' and the 
related linear and non linear optical effects~\cite{baba}. A new exciting area in 
photonics is thus very much on the cards.

Inspired by these experiments and other works on localization of 
light in tailor made lattices, 
in the present communication we address the problem of localization of electromagnetic waves 
and a controlled engineering of the flat, non dispersive photonic bands in 
a kagom\'{e} strip, made of single mode waveguides. 
In this work each elementary unit cell of the kagom\'{e} geometry encloses  
a Sierpinski gasket (SPG)~\cite{kadanoff} network made of similar single mode waveguides, and 
of arbitrarily large generation (~Fig.\ref{kagome}(a)). This construction introduces 
a non trivial modification of the existing waveguide networks 
in the kagom\'{e} class, and throws an 
achievable challenge to the experimentalists in this era of pretty advanced 
nanotechnology and lithography techniques. 

The principal motivation behind this idea is to look into the possibility of generating a 
whole hierarchy of non-dispersive, flat optical bands, if they exist, and to 
work out a scheme for locating them in the optical spectrum of the 
kagom\'{e}-fractal system. An SPG 
network, in the context of electrons, is already known to possess an infinite 
number of extended, totally transmitting eigenstates even in the absence of 
any long range translational order. 
Such states densely populate the otherwise Cantor set energy spectrum consisting of 
localized eigenfunctions~\cite{ac3}.
This is likely to be the case for optical 
modes as well. If it is true, then the proximity 
of sharply localized non-dispersive modes to the totally transmitting (extended), 
dispersive ones (as the fractal grows in size)  
may lead to a new interesting scenario where a comprehensive control 
can be achieved over a possible {\it transition} between {\it slow} and {\it resonant} 
optical modes as a function of the frequency of the light injected into the waveguide 
network.
   
It should also be appreciated that, while a kagom\'{e} geometry in its simplest 
shape, is already known to exhibit flat 
photonic bands, the incorporation of the fractal geometry in each unit cell destroys the 
translational order locally (though it is preserved on a global scale in the horizontal 
direction) in the transverse direction. This makes the conventional methods of extracting 
flat band states impossible to be implemented, specially as the gasket grows in size. 
Straightforward diagonalization of hamiltonian (for
electronic case) or an exact numerical solution of the homogeneous wave equation
may not help because of highly fragmented spectral character, and the eigenmodes
obtained from a finite size system may slip away from the original spectrum as we
move over to a higher generation.  
We prescribe a simple method to circumvent this problem 
by exploiting the scale invariance of the gasket structure 
and a real space renormalization group (RSRG) method. The method can be applied 
practically to any quasi-one dimensional systems and, to the best of our knowledge, 
has not been reported before as far as the evaluation of flat photonic bands are 
concerned.

The other point of interest in the present work is to analyze the distribution of 
the amplitudes of such flat band states in real space. Recently, a class of localized 
eigenfunctions in deterministic fractal geometries has been revealed both in the 
context of electronic states and optical modes~\cite{biplab2,biplab3} which can 
in principle, be {\it delayed} in space by an appropriate choice of the 
energy of the electron or the frequency of the injected wave. Such states exhibit 
a fragmented distribution in the respective spectra, a consequence of the underlying 
lattice topology. We would like to examine whether such states, if they exist in 
the present network, can be discerned inspite of the complicated geometry of the network, 
and that, if they really correspond to the non dispersive, flat character which has been 
of immense interest recently. 

Before we end this section, 
it is worth mentioning that,
single mode waveguides have recently been fabricated 
and used in a quasiperiodic optical setup~\cite{kraus} 
to unravel topological states in quasicrystals. Localization transition in 
one dimensional quasiperiodic lattices has also been investigated experimentally in recent 
times~\cite{lahini}. The present proposal, to our mind, can thus be tested with an appropriately 
designed waveguide network.

In what follows we describe our findings. In section~\ref{sec2} we describe 
the wave equation and the RSRG scheme. Section~\ref{results} contains a discussion
regarding the density of eigenmodes, existence of non-dispersive eigenmodes, 
and the dispersion
relation. 
Finally, in section~\ref{conclu} we draw our conclusions. 
 
\section{The wave equation and the RSRG scheme}
\label{sec2}
\subsection{The difference equations}

\begin{figure}[ht]
\centering
\includegraphics[clip,width=8.5cm,angle=0]{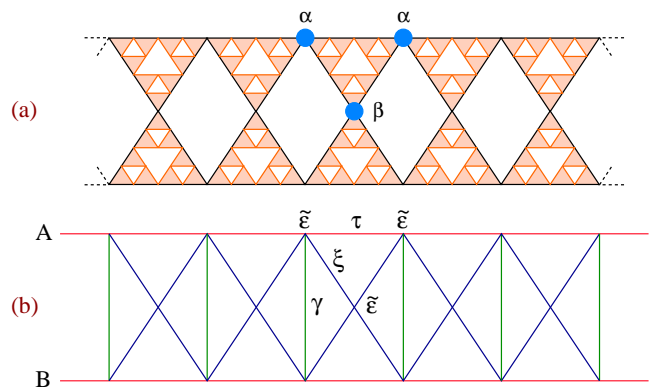}
\caption{(Color online) (a) An infinite fractal incorporated kagom\'{e} waveguide
 network and (b)
the effective \textit{renormalized} ladder network.}  
\label{kagome}
\end{figure}

Let us begin by refering to Fig.~\ref{kagome}(a), where we have taken a typical waveguide network
formed by waveguide segments having the same dimensions arranged 
in a kagom\'{e} geometry.
A Sierpinski gasket fractal network of finite generation is embedded inside the
skeleton of kagom\'{e} geometry within each `unit cell'. Each segment has a
single channel for wave propagation. The wave propagation inside the core of the waveguide
is governed by the wave equation~\cite{alex,sheng}, viz.,
\begin{equation}
\dfrac{\partial^2\psi_{ij}(x)}{\partial x^2}+k^2\psi_{ij}(x)=0
\label{waveeqn}
\end{equation}
where, $\psi_{ij}(x)$ is the photonic wavefunction between two nodes $i$ and $j$,
and $k=\frac{\omega \sqrt{\epsilon_r}}{c_0}$ is the wave vector inside the material of
the waveguide. $\omega$ is the frequency of the wave propagating, $c_0$ is the speed
of light in free space and $\epsilon_r$ contains all the dielectric information of the
material of the core of the waveguide. The above Eq.~\eqref{waveeqn} has the
obvious solution of the form ~\cite{alex,sheng},
\begin{equation}
\psi_{ij}(x)=\psi_i \dfrac{\sin[k(l_{ij}-x)]}{\sin kl_{ij}}+\psi_j
\dfrac{\sin kx}{\sin kl_{ij}}
\label{wavesoln}
\end{equation}
$x$ is the variable distance measured from $i$-th nodal point, $l_{ij}$ is the length
of the waveguide segment between $i$-th and $j$-th node and $\psi_i$ and $\psi_j$ 
characterizes the wavefunction at the corresponding nodes respectively. The flux 
conservation condition yields,
\begin{equation}
\sum_{j}\left[\dfrac{\partial}{\partial x} \psi_{ij}(x)\right]=0
\label{fluxconserv}
\end{equation}
where the summation $j$ runs over all possible nodes connected directly to $i$, and 
this finally
leads to a discretized version of Eq.~\eqref{waveeqn}, viz.,
\begin{equation}
-\psi_i\sum_{j}\cot\theta_{ij}+\sum_{j}\psi_j/\sin\theta_{ij}=0
\label{difference1}
\end{equation}
where $\theta_{ij}=kl_{ij}=ka$, $a$ being the constant waveguide dimension. The above
Eq.~\eqref{difference1} is an exact replica of the equation for 
spinless non-interacting
electrons propagating in a lattice.
The difference equation, an equivalent discretized form of Schr\"{o}dinger's equation in this 
latter case reads,
\begin{equation}
(E-\epsilon_{i}) \psi_{i} = \sum_{j}t_{ij} \psi_{j}
\label{difference2}
\end{equation}
Here, $E$ is the electron energy, $\epsilon_i$ is the onsite potential at each atomic site,  
and $t_{ij}$ , the nearest neighbor hopping integral, signifying the overlap
of electronic wavefunctions between two neighboring sites.

We exploit this resemblance to analyse the existence of dispersionless flat photonic
bands as well as the extended Bloch-like eigenmodes in such typical waveguide system.
It is trivial to observe that, a replacement of $ka$ by $\pi-ka$ yields new recast version
of Eq.~\eqref{difference2} with $t_{ij}$ being replaced 
by $-t_{ij}$ which does not
affect the spectrum in any way.

The one-to-one correspondence of Eq.~\eqref{difference1} and Eq.~\eqref{difference2},
prompts us to 
identify $E - \epsilon_i =-\sum_{j}\cot\theta_{ij}$ and $t=1/\sin ka$. It is then 
simple to extract information about the photonic modes using the language of the 
electrons, and the results will be exact.
As illustrated in 
Fig.~\ref{kagome}, the fractal-kagom\'{e} waveguide network contains 
nodes with four nearest neighbors only. Therefore, for classical wave propagation
we just assign $E - \epsilon_i=-4\cot ka$ for every vertex `$i$'. The {\it overlap integral} along the
arm of each basic triangular plaquette is $t=1/\sin ka$ carries the kinetic information
of the propagating wave.

Using a real space renormalization group method, we now proceed to describe 
our basic scheme and the physics of wave propagation 
in such fractal incorporated kagom\'{e} geometry by
exploiting the exact analogy with the corresponding electronic problem.
\subsection{Converting an SPG-kagom\'{e} network into a two-strand 
ladder}

Let us suppose that an SPG at its $n$th generation is grafted inside every unit cell 
of the kagom\'{e} strip (Fig.~\ref{kagome}(a)). 
As proposed before, we set $\epsilon_i = E + 4 \cot ka = \epsilon$ at each node and 
$t = 1/\sin ka$ ($ka =\theta_{ij}$ is 
assumed constant,  
as we take all the waveguide segments to be identical in every respect). $E$ can be 
chosen quite arbitrarily (playing the role of Fermi energy in the equivalent 
electronic case), and this selection in the present case sets the range of frequency  
of the injected wave. 

With this assumption,
we decimate out a subset of the wave amplitudes
on the bulk vertices of the SPG fractal system in terms of others surviving vertices to
generate an  
effective `two-arm ladder' (Fig.~\ref{kagome}(b)). This ladder is now 
characterized by the renormalized values of the on-site potentials at its vertices, 
$\tilde\epsilon = \epsilon_n + 2 t_n^2/(E-\epsilon_n)$, 
the hopping integral along each arm $\tau = t_n + t_n^2/(E-\epsilon_n)$, 
the inter-arm (between arms $A$ and $B$)
hopping integral $\gamma = 2 t_n^2/(E-\epsilon_n)$, and 
the second neighbor hopping inside every plaquette $\xi = t_n^2/(E-\epsilon_n)$.
This latter quantity is a result of the decimation of the `internal' vertices in a unit cell.

The quantities $\epsilon_n$, and $t_n$ are obtained following the usual RSRG 
recursion relations in the analogous electronic case~\cite{ac3}, viz.,
\begin{equation}
\begin{aligned}
& \epsilon_{n} = \epsilon_{n-1} + 
\dfrac{4t_{n-1}^{2} (E-\epsilon_{n-1})}
{(E-\epsilon_{n-1})^{2} - 
(E-\epsilon_{n-1})t_{n-1} 
- 2t_{n-1}^{2}} \\
& t_{n} = \dfrac{2t_{n-1}^{3} + 
(E-\epsilon_{n-1})t_{n-1}^{2}}
{(E-\epsilon_{n-1})^{2} - 
(E-\epsilon_{n-1})t_{n-1} 
- 2t_{n-1}^{2}}
\end{aligned} 
\label{recursion}
\end{equation}
with $\epsilon_0 = \epsilon$ and $t_0 = t$.
The difference equations for the ladder network now read,
\begin{widetext}
\begin{eqnarray}
(E - \tilde{\epsilon}) \psi_{n,A} & = & \tau (\psi_{n+1,A} + \psi_{n-1,A}) + 
\gamma \psi_{n,B} + \xi (\psi_{n+1,B} + \psi_{n-1,B}) \nonumber \\
(E - \tilde{\epsilon}) \psi_{n,B} & = & \tau (\psi_{n+1,B} + \psi_{n-1,B}) + 
\gamma \psi_{n,A} + \xi (\psi_{n+1,A} + \psi_{n-1,A}) 
\label{difflad}
\end{eqnarray}
\end{widetext}

The Eq.~\eqref{difflad} can be written in an equivalent matrix form, viz., 
\begin{widetext}
\begin{eqnarray}
\left [
\left( \begin{array}{cccc}
E & 0 \\
0 & E
\end{array}
\right ) - 
\left( \begin{array}{cccc}
\tilde\epsilon & \gamma \\
\gamma & \tilde\epsilon
\end{array}
\right)
\right ]
\left ( \begin{array}{c}
\psi_{n,A} \\
\psi_{n,B}  
\end{array} \right )
& = & 
\left( \begin{array}{cccc}
\tau & \xi \\ 
\xi & \tau 
\end{array} 
\right)
\left ( \begin{array}{c}
\psi_{n+1,A} \\
\psi_{n+1,B}  
\end{array} \right )
+
\left( \begin{array}{cccc}
\tau & \xi \\                                                  
\xi & \tau
\end{array}
\right)
\left ( \begin{array}{c}
\psi_{n-1,A} \\
\psi_{n-1,B}
\end{array} \right )
\label{eqladder1}
\end{eqnarray}
\end{widetext}
The $2 \times 2$ `potential matrix' involving 
$\tilde\epsilon$ and $\gamma$, and the `hopping matrix' with matrix elements 
$\tau$ and $\xi$ commute. This means they can be diagonalized simultaneously 
using a similarity
transform. Taking advantage of this, we make a uniform change of basis, going from 
the basis $\mathbf{\Psi_n \equiv (\psi_{n,A}, \psi_{n,B})}$ to 
a {\it new} basis $\mathbf{\Phi_{n} \equiv (\phi_{n,A}, \phi_{n,B})}$, such that,  
$\mathbf{\Phi_{n}}=\mathbf{M}^{-1}\mathbf{\Psi_n}$.  
The matrix $\mathbf{M}$
diagonalizes both the
`potential' and the `hopping' matrices. This leads to a decoupling of the 
coupled set of equations Eq.~\eqref{difflad}~\cite{sil1,sil2}. 
The decoupled equantions read, 

\begin{widetext}
\begin{eqnarray}
\left [ E - \left (\epsilon_n - \frac{4t_n^2}{E-\epsilon_n} \right ) \right ] 
\phi_{n,A} & = & \left(t_n+\frac{2t_n^2}{E-\epsilon_n}\right) 
( \phi_{n+1,A} + \phi_{n-1,A} ) \nonumber \\
( E - \epsilon_n) \phi_{n,B} & = & t_n ( \phi_{n+1,B} + \phi_{n-1,B} )
\label{decouple1}
\end{eqnarray}
\end{widetext}
The decoupled equations individually represent two independent linear chains 
along which some quantum particle propagates, and its spectrum will be 
identical to that of the photonic waveguide we have constructed, as the 
equivalence between the two cases is an exact one. 

We now proceed to explore these equations to unravel the  
dispersionless modes, if any, and the others for the kagom\'{e} fractal waveguide.
\section{Results and discussion}
\label{results}
\subsection{The general character of the spectrum}
Before looking for the flat photonic bands and the extended eigenmodes, it is advisable to have a 
general idea of the spectrum of allowed photon frequencies in such a system.
It should be appreciated that, each of the decoupled equations in 
Eq.~\eqref{difflad} represents a purely ordered chain of identical `atomic' sites 
with effective, energy dependent potentials and hopping integrals. Each equation 
has its own density of states spectrum, and a convolution of these will give the 
true density of states of the actual system. The fragmentation of the spectra and 
their subband structures will depend on the generation of the embedded SPG network.

We have however, used an RSRG decimation scheme on the kagom\'{e} fractal  
in Fig.~\ref{kagome}(a) to work out the local Green's function $G_{00} = G_{00}^{\alpha} + 
G_{00}^{\beta}$. The superscripts $\alpha$ and $\beta$ refer to the top and middle vertices 
in the kagom\'{e} skeleton, as marked by solid blue circles in Fig.~\ref{kagome}(a). 
We refrain from including the detailed mathematical expressions exploited to arrive at the 
result just to save space, and provide a typical 
\begin{figure}[ht]
\centering
\includegraphics[clip,width=8cm,angle=0]{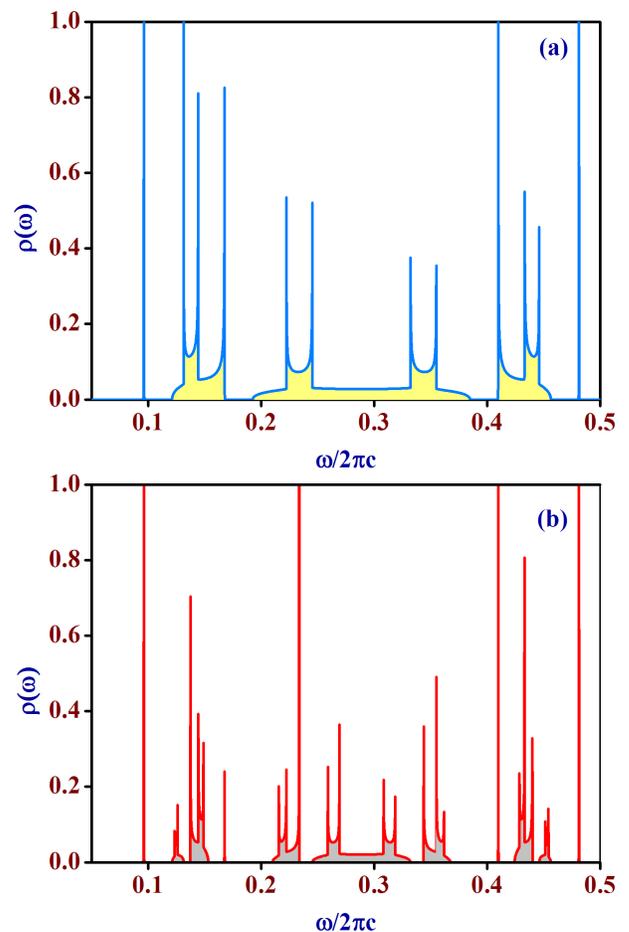}
\caption{(Color online) Density of eigenmodes as a function of normalized 
frequency for a fractal-kagom\'{e} waveguide network when an SPG geometry of (a)
$1^{st}$ generation and (b) $2^{nd}$ generation is embedded into it.}  
\label{newados2}
\end{figure}
local density of photonic modes (LDPM) as a
function of normalized frequency when first and second generation SPG fractals 
are inserted in an infinite kagom\'{e} strip is shown in  
Fig.~\ref{newados2}, as obtained from the well known expression,
\begin{equation}
\rho(\omega)=(-1/\pi) \lim_{\eta \rightarrow 0}\text{Im}\;G_{00} (\omega+i\eta)
\end{equation}

The distribution of eigenmodes plotted within $0<\omega/2\pi c<1/2$ shows clusters of nonzero
values of LDPM over finite range of $\omega$. Quite arbitrarily we have set $\epsilon_r=3$. 
The network
exhibits gaps in the frequency spectrum in which the propagation of wave
is not allowed and thus is a model PBG system. With increasing generations of the 
embedded SPG fractal the LDPM gradually assumes a fragmented Cantor-like character, 
the key signature of 
the fractal itself that fills up a kagom\'{e} unit cell. The signature is already 
apparent in Fig.~\ref{newados2}. 

\subsection{The Non-dispersive eigenmodes}
If a single photonic mode is dispersionless in one direction, it implies 
localization in
that direction as an extreme case, unfolding the possibility of 
engineering ultra slow light. This results from a 
perfect geometric phase cancellation which causes an element of the 
{\it effective mass tensor} to be divergent, leading to a possible 
`observation' of {\it heavy photons}. 
This doesn't look improbable if we borrow the language for the 
electrons, thanks to the analogous tight binding model which indeed works quite 
accurately for the low lying photonic dispersion. Flatness in the 
frequency ($\omega$)-vs-wave vector ($q$) curves implies 
infinite {\it effective mass} of the particle involved.
This in turn, makes the group velocity vanish. The kinetic energy of the 
associates wave packet is quenched, and 
the density of states diverges due the relationship 
\begin{equation}
\rho_{\omega} \propto \int v_{g}^{\:-1} d^{3}q
\end{equation}
where, $v_{g} \propto d\omega/dq$~\cite{apaja}. Such singularities in
the density of states are expected to produce anomalous behaviors in physical
properties as well as transport phenomena and optical response. 

Because of zero group
velocity, the corresponding single particle state is sharply localized
at a point, or in a finite cluster of nodal points in the system. Such clusters are
separated from the neighboring clusters by vertices where the amplitude of the
wave function is zero. Therefore, the movement of the particle is restricted along
the periphery of those finite clusters. 
The important question is, how can we extract such eigenvalues for 
a kagom\'{e}-fractal network?

In the first of decoupled set of equations Eq.~\eqref{difflad}, if we 
set $t_n+2 t_n^2/(E-\epsilon_n) = 0$, we get
$E=\epsilon_n-2 t_n$.  The first equation now represents, in the 
language of the tight binding scheme, the difference 
equation of an isolated
(effective) `atom'. The corresponding eigenstate(s) will be sharply localized, 
provided one gets non-zero density of states at these special energies. 
These will be the desired non dispersive, flat band (FB) modes, as will be 
explicit towards the end of this subsection.
With 
$\epsilon_r=3$, the frequencies of the `flat band mode' turn out to be 
$\omega/2\pi c = 0.096225$, $0.167561$, $0.409795$, and $0.48113$. These correspond to 
the isolated spike at the left end of Fig.~\ref{newados2}(a), the right and the 
left `edges' of 
the first and third subbands (yellow shaded) respectively, 
and the sharply localized mode at extreme
right. The roots corresponding to the edges ($0.167561$ and $0.409795$) 
separate out of the bands when the second 
generation SPG is inserted into the unit cell of the kagom\'{e} strip, as seen 
from Fig.~\ref{newados2}(b). Obviously, 
as we extract the roots from the deeper and deeper scales
 of length (which means sequentially setting $n=1$, $2$, $3$, $.....$), 
more and more such dispersionless
FB states result. Every FB mode arising out of a given generation $n$ and 
forming the edge of a subband gets detached  
from it and stand out separate for all subsequent generations $n+m$, with $m=1$, $2$, 
$3$, $.......$.

For an electromagnetic wave with an arbitrary wave vector $q$ 
injected in the SPG-kagom\'{e} network,  
each of the two decoupled equations Eq.~\eqref{difflad}, 
can be simultaneously used in analogy with the dynamics of an electron in a periodic
\begin{figure}[ht]
\centering
\includegraphics[clip,width=8cm,angle=0]{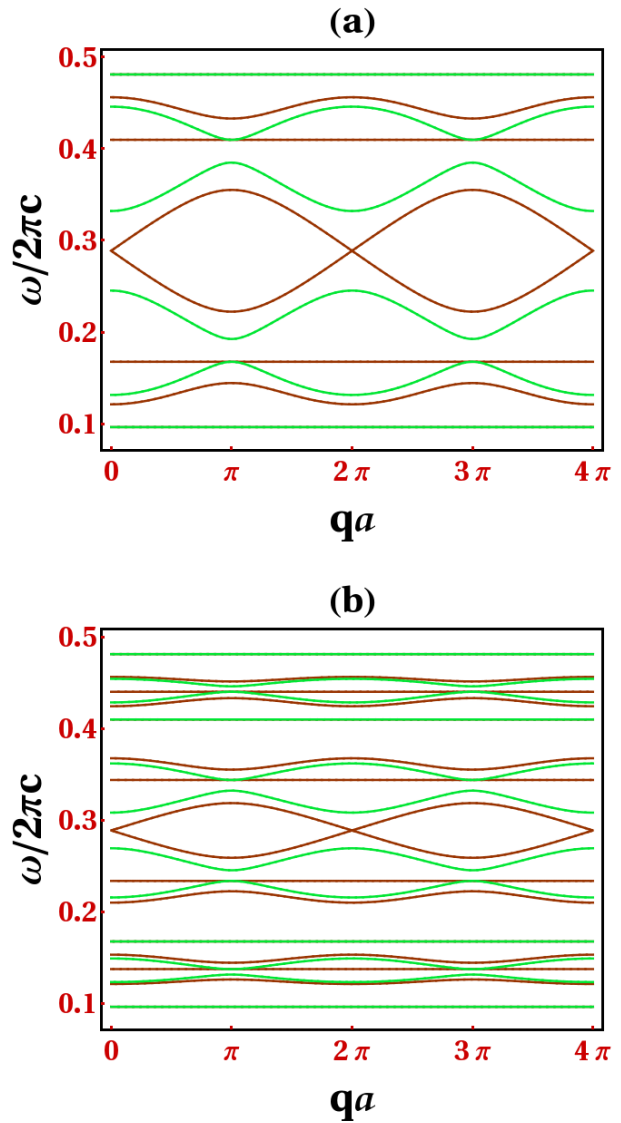}
\caption{(Color online) Dispersion relation for a fractal-kagom\'{e} waveguide network
 when a $b=2$ SPG geometry of (a)
$1^{st}$ generation and (b) $2^{nd}$ generation is embedded into it.
The green and brown lines correspond to solutions obtained from the first 
and the second decoupled equations respectively.}  
\label{disp}
\end{figure}
array of (effective) potentials, to obtain the dispersion relation for photons.
This obviously displays the $\omega/2\pi c - qa$ curves, bringing out the 
`allowed' photonic bands as functions of the injected wave vector.
The  
photonic dispersion relations, for materials with $\epsilon_r=3$, and when 
a $1^{st}$ generation SPG is inserted in the kagom\'{e} motif read,
\begin{widetext}
\begin{eqnarray}
\left [2 \cos(2\pi\sqrt{\epsilon_{r}}\omega' a) - 1 \right ]
 \left [4 \cos(2\pi\sqrt{\epsilon_{r}}\omega' a) + 1 \right ]
\left [6 \cos(2\pi\sqrt{\epsilon_{r}}\omega' a) + 4 \cos(4\pi\sqrt{\epsilon_{r}}\omega' a)
 + 3 - \cos qa \right ] & = & 0 \nonumber \\
\left [6 \cos(2\pi\sqrt{\epsilon_{r}}\omega' a) + 4 \cos(4\pi\sqrt{\epsilon_{r}}\omega' a)
 + 4 - \cos qa \right ] 
\left (2 \cos(2\pi\sqrt{\epsilon_{r}}\omega' a) - 1 \right) & = & 0 
\label{disp2}
\end{eqnarray}
\end{widetext}
In the above Eq.~\eqref{disp2} $\omega'=\omega/2\pi c$.
The flat, dispersionless modes are easily extracted by setting  
$[4 \cos (2\pi \sqrt{\epsilon}\omega' a) + 1] = 0 $, and, 
$[2 \cos (2\pi \sqrt{\epsilon}\omega' a) - 1] = 0$ respectively. The roots  
match exactly with the values of the normalized frequency $\omega/2\pi c$ for the 
sharply localized, {\it atomic like} states, as evaluated earlier.

In the top panel of Fig.~\ref{disp}(a) we show the roots of Eq.~\eqref{disp2} 
when a $1^{st}$ generation SPG is embedded in the unit cells of a kagom\'{e} strip. Completely 
flat frequency ($\omega/2\pi c$) versus wave vector ($q$) are apparent. The lower panel exhibits 
similar flat bands, now more in number as a higher ($2^{nd}$) generation SPG network is grafted 
in the kagom\'{e} strip.

It is interesting to work out that the roots obtained 
as solutions of the equation $E - \epsilon_n + 2t_n = 0$ are also solutions of the 
equations $E - \epsilon_{n+1} + 2t_{n+1} = 0$. That is, the flat band eigenmodes
exhibit a nested pattern. This is also explicitly seen from the figures displayed 
in Fig.~\ref{disp}. Interestingly, the outermost flat bands remain `undisturbed', 
while the neighborhood of the inner ones get densely packed with dispersive and 
non dispersive modes as one increases the generation of the embedded SPG. 

\subsection{The extended eigenmodes and the crossover}
The similarity of the wave equation and the tight binding one for the electrons help us 
extract an infinite number of {\it extended} Bloch-like eigenmodes in the kagom\'{e}-fractal 
network. This is simply done as one observes, using Eq.~\eqref{recursion} that once we fix 
$E = \epsilon_n$ at any $n$-th stage of renormalization, the `hopping integral' gets 
locked into a $2$-cycle fixed point, viz., $t_{n+2}=-t_{n+1}=t_n$. A non-zero value 
of the hopping integral at all stages of renormalization implies finite overlap of the 
`wavefunction' at all scales of length - a clear signature of the corresponding mode 
being extended (a transparent state, in terms of transmission). For example, if we set 
$E=\epsilon_1$, the principal values of the resulting frequencies 
corresponding to the extended photonic mode for waveguide 
materials with $\epsilon_r=3$ turn out to be $\omega/2\pi c=1/4\sqrt{3}$, 
$\sqrt{3}/4$.
Out of these, the first root is a solution of the equation $E=\epsilon_0$ 
at the bare length scale. 
The Eq.~\eqref{recursion} exhibits a nesting pattern of the roots in the sense that, the solutions 
of the equation $E-\epsilon_n=0$ includes the solutions already obtained from the equation 
$E-\epsilon_{n-1}=0$. 

As one should now be able to appreciate that, as one inserts larger and larger versions of the 
SPG inside a kagom\'{e} unit cell, both the flat band eigenmodes and the extended ones get 
densly packed in the dispersion curves (as well as in the LDPM graphs). 
The curvatures of even the dispersive modes start becoming lesser and lesser indicating 
a lowering of the group velocity. The dense packing of the flat and the dispersive 
bands therefore opens up a possibility of engineering the so called {\it slow light} 
and re-entrant {\it crossover} of optical eigenmodes, going from an extended (dispersive) 
to a sharply localized (flat) one, or vice versa as one climbs the frequency axis 
at any fixed value of the wave vector $q$. Such a crossover can, in principle, be 
controlled over arbitrarily small intervals of energy, thanks to the intrinsic multifractal 
character of the energy spectrum of the embedded SPG geometry.
\subsection{Spatial distribution of the flat band states and the staggered localization 
effect}

\begin{figure}[ht]
\centering
\includegraphics[clip,width=8cm,angle=0]{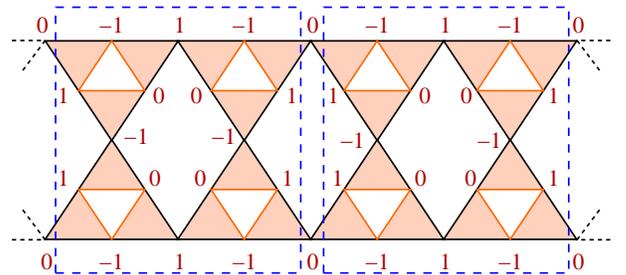}
\caption{(Color online) Distribution of amplitude for the flat non-dispersive photonic
 state at
$\omega=0.096225$ (in the unit of $2\pi c$) for the fractal incorporated 
kagom\'{e} waveguide network. 
The blue dotted zones are the \textit{allowed} zones getting decoupled from each other
 by the vertices having zero amplitude of the wavefunctions.}   
\label{amplitude}
\end{figure}

The localized character of the eigenmodes are illustrated in Fig.~\ref{amplitude}, where
the amplitudes of the photonic modes are shown only at the vertices of the 
kagom\'{e} motif for a clearer understanding. The `spanning' plaquettes, in which the 
amplitudes are mostly confined are shaded for identification. Fig.~\ref{amplitude} 
brings out distribution corresponding to the basic flat band frequency $\omega/2\pi c=0.096225$. 
One blue
dotted square is `separated' from its neighboring squares by a couple of sites on which 
the amplitudes disappear. For a frequency extracted from a higher generation $n$, the 
amplitudes spread over larger clusters, but one such cluster remains {\it decoupled} 
from its neighbors through vertices at which the amplitude of the excitation vanishes.

It is encouraging to extend this concept when increasing generations of the 
SPG occupy the plaquettes. One can then extract flat, non-dispersive modes for 
which the distribution of amplitudes will assume a non trivial shape over larger 
clusters of vertices. Thus, the {\it onset of localization} will be felt at a relatively 
larger scale of length. How large - depends on the scale (or generation index $n$) 
at which we are solving the dispersion equation.
 
This leads to the concept of \textit{staggered localized} modes, where one
can have the liberty to delay the onset of localization {\it in space} using the
RSRG index, and an appropriate frequency~\cite{biplab3}. 

The FB states talked about so far are in fact localized around each crossing,
with a rapidly decaying tail toward the neighboring junctions.
The difference equation corresponding to each vertex can easily 
be exploited to construct a bound state amplitude
configuration such as shown in the Fig.~\ref{amplitude}. 
By contrast, resonant states or the dispersive states, which have
intrinsic losses, will couple to extended states to result in
significant transfers to distant neighbors, deteriorating the
tight-binding picture with near-neighbor transfers.

\section{Closing Remarks}
\label{conclu}
In conclusion, we have critically examined the photonic spectral structure a the kagom\'{e}
monomode waveguide strip which with embedded Sierpinski gasket fractal mesh in every unit cell.
The spectrum is shown to contain a countable infinity of flat, 
dispersionless optical modes which are localized in clusters of finite
size, displaying a multitude of localization area. Such clusters  
are \textit{effectively} separated from similar clusters in the
network by nodes where the amplitude of the wave function
is zero. We have been able to present an analytical prescription for the determination
of frequency of such non-dispersive eigenmodes. The size of the clusters can be tuned
at will by the length scale at which the frequency is evaluated. The onset of localization
may be delayed in space (\textit{staggered}). 

Such results
give an opportunity to modulate experimentally the
localization of classical waves, for example, light, triggered by
the lattice topology without bothering about the high permittivity of the core 
materials. It might be useful in developing novel
photonic band-gap structures, where the {\it pass band} can be controlled over
arbitrarily small domains by exploiting the fractal signature of the network.
Also, the physical significance of vanishing group velocity is that several scattering
waves form standing
wave pattern in such photonic waveguide network. For such \textit{self-localized}
 eigenmodes, we can
obtain coherent waves, i.e., lasing action at the photonic band edges.
\begin{acknowledgements}
AN acknowledges financial support through a research fellowship
[Award letter no. F.$17$-$81 / 2008$ (SA-I)] from UGC, India. AC acknowledges a partial 
financial support from University of Kalyani through its DST-PURSE grant.
\end{acknowledgements}

\end{document}